\documentstyle[12pt]{article}
\date{}

\newcommand\de{\delta}
\newcommand\eps{\epsilon}

\newcommand\lam{\lambda}

\newcommand{\dl}{{\bf L}}

\newcommand{\del}{{\bf \nabla}}

\newcommand\btd{\raise 2pt
\hbox{$\hat\bigtriangledown$}\hskip 1.5pt}
\newcommand\bt{\raise 2pt
\hbox{$\bigtriangledown$}\hskip 1.5pt}
\begin{document}
\baselineskip=14 pt

\begin{flushright}
January, 2002
\end{flushright}
\vspace{0.1in}
\begin{center}{\large\bf Black Hole Mass Formula  \\[2mm]
Is a Vanishing Noether Charge}
\end{center}

\begin{center}
Xiaoning Wu$^a$,\quad Han-Ying Guo$^a$,\quad Chao-Guang
Huang$^{b,a}$\quad and \quad Ke Wu$^{c,a}$
\end{center}
\begin{center}{\small
a. Institute of Theoretical Physics, Chinese Academy of Sciences,\\
\quad  P.O. Box 2735, Beijing 100080, China.\\
b. Institute of High Energy Physics, Chinese Academy of
Sciences,\\
\quad P.O. Box 918(4), Beijing 100039, China.\\
c. Department of Mathematics, Capital Normal University,\\
Beijing 100037, China.}\end{center}
\par\
\begin{abstract}
 The Noether current and its variation relation
with respect to diffeomorphism invariance of
gravitational theories have been derived from the
horizontal variation and vertical-horizontal
bi-variation of the Lagrangian, respectively. For
Einstein's GR in the stationary, axisymmetric
black holes, the mass formula in vacuum can be
derived from this Noether current although it
definitely vanishes. This indicates that the mass
formula of black holes is a vanishing Noether
charge in this case. The first law of black hole
thermodynamics can also be derived from the
variation relation of this vanishing Noether
current.
\end {abstract}
 \vskip 10mm

{\bf 1.} It is well known that Noether's theorem
links the conservation law of certain quantity
with certain symmetry. From 1970's, one of the
most exciting discoveries in GR is the black hole
mechanics that relates among others the geometric
properties of black holes with the thermodynamic
laws. In thermodynamics, the first law  exhibits
the relation among the changes of energy, entropy
and other macroscopical quantities. On the other
hand, in either classical mechanics or field
theories the energy conservation law is directly
related with the time translation invariance.
Furthermore, its covariant form is closely
related with the re-parameterization invariance
of spacetime coordinates, i.e. the diffeomorphism
invariance of the theory {\footnote{Generally
speaking, diffeomorphism is not a group but a
pseudo-group. In some literatures, the term of
diffeomorphism invariance is restricted to the
re-parameterization invariance of spacetime
coordinates that keeps the line-element $ds^2$
being invariant. This restriction is adopted
here.}}. Therefore, it is significant to explore
whether the diffeomorphism invariance of the
gravitational theories should be behind the mass
formula and its differential one, i.e. the first
law of black hole thermodynamics.

During last decade, Wald and his collaborators as
well as other authors (see, for example,
\cite{wald1}-\cite{wald5}) have studied this
problem and found certain link between the first
law of black hole thermodynamics and the
diffeomorphism invariance of the gravitational
theories. They have also claimed that the black
hole entropy is a Noether charge
\cite{wald2},\cite{wald3},\cite{wn}. But, some
problems are still open even for the classical
black hole thermodynamics such
as how to relate the formula for mass %
itself with the conservation current with respect
to the diffeomorphism invariance, in what sense
the black hole entropy is a Noether charge and so
on so forth.

In this letter we study these problems by the
variational approach. We mainly work out the
relation between diffeomorphism invariance and
the mass formula as well as the first law of
black hole thermodynamics in a manifest way. We
first sketch the derivation for the Noether
current and its variation relation with respect
to the diffeomorphism invariance of the
Lagrangian 4-form by taking horizontal variation
and vertical-horizontal bi-variation of the
Lagrangian 4-form  of the gravitational
theories on $({\cal M}^4, g)$, respectively. %
For the vacuum gravitational fields in GR, we
show that this current vanishes definitely.
However, for the vacuum stationary, axisymmetric
black holes,  this vanishing Noether charge of
the vanishing current leads to the formula for
mass. This indicates that the entropy of the
black hole itself cannot be regarded as a Noether
charge at least for the vacuum cases rather the
mass formula, as a whole, should be regarded as
the vanishing Noether charge in certain sense.
Further, we derive the first law of black hole
thermodynamics \cite{sm} from the variation
relation of the Noether current. Finally, we end
with some conclusion remarks. 

\vskip 4mm {\bf 2.} Let ${\bf L}$ be a
diffeomorphism invariant Lagrangian 4-form for
gravitational field metric theory on the
space-time manifold $({\cal M}^4, g)$ with metric
$g$ of signature $-2$. This means that if $f$ is
a diffeomorphism map on ${\cal M}^4$, the
Lagrangian 4-form satisfies
\begin{eqnarray}
f^*({\bf L}(g_{ab}))={\bf L}(f^*(g_{ab})),
\end{eqnarray}
where  $f^*$ is the induced map by $f$.

The horizontal variation of ${\bf L}$ with
respect to the diffeomorphism  is defined as
\begin{eqnarray}
\hat\de{\bf L}=\frac{d}{d\lam}{\bf
L}(f^*_{\lam}g_{ab})|_{\lam=0}={\cal L}_\xi{\bf L}(g_{ab}),
\end{eqnarray}
where $f^*_{\lam}$ denotes the induced map of a
one parameter diffeomorphism group $f_{\lam}$
generated by a vector field $\xi^a$ on ${\cal
M}^4$, ${\cal L}_{\xi}$ is the Lie derivative
with respect to $\xi$.

It is easy to show the following formal equation
(see, for example, \cite{wald1})
\begin{eqnarray}
 \label{f} \hat\de{\bf L}={\bf E}\hat\de g+d{\bf
\Theta} (\hat\de g),
\end{eqnarray}
where %
${\bf E}=0$ gives rise to the Euler-Lagrange
equation for the gravitational field theory and
${\bf \Theta}$ the symplectic potential. In
addition, from the diffeomorphism invariance of
the Lagrangian it follows
\begin{eqnarray}
\label{li} \hat\de\dl={\cal L}_{\xi}\dl
=d(\xi\cdot\dl).
\end{eqnarray}
Combination of eqs.(\ref{f}) and (\ref{li})
directly leads to a conservation equation
\begin{eqnarray}\label{Neq}
d*{\bf j} +{\bf E}\hat\de g=0,
\end{eqnarray}
where $\bf j$ is the Noether current with respect
to the diffeomorphism invariance
\begin{eqnarray}\label{cc}
{\bf j}:=*({\bf \Theta}(\hat\de g)-\xi\cdot\dl),
\quad mod(*d{\mbox {\boldmath $\alpha$}}).
\end{eqnarray}
It is conserved if and only if the field equation
${\bf E}=0$ holds. This current (\ref{cc})
derived from the horizontal variation of the
Lagrangian 4-form with respect to the
diffeomorphism invariance is in the same form as
the one defined by Wald et al. Note that if $\xi$
is a Killing vector, eq. (\ref{li}) vanishes and
the Noether current becomes
\begin{eqnarray}\label{cck}
{\bf j}:=*{\bf \Theta}(\hat\de g), \quad
mod(*d{\mbox {\boldmath $\beta$}}).
\end{eqnarray}

In order to derive the differential formula of
mass, i.e. the first law of the black hole
thermodynamics, it is needed the variation
relation of this Noether current. To this end, it
is natural to calculate the following
bi-variation with respect to vertical and
horizontal variation $\delta$ and $\hat\delta$ of
${\bf L}$,
\begin{eqnarray}
\label{bi} \delta\hat\delta {\bf L} = \delta({\bf
E}\hat\de g+d{\bf \Theta} (\hat\de
g))%
=\delta d(\xi\cdot\dl).%
\end{eqnarray}
Therefore, from eqs. (\ref{bi}) and (\ref{cc}),
it follows the variation relation of this Noether
current
\begin{eqnarray}\label{dcc}
0=\delta({\bf E}\hat\de g)+ d{\delta *\bf j},
\end{eqnarray}
where it has been used the commutative property
between the vertical variation operator and the
differential operator, i.e. $\de d=d\de$. Thus,
for the conservation of  variation of the Noether
current (\ref{cc}), i.e.
\begin{eqnarray}\label{dcc1}
d{\delta *\bf j}=0,
\end{eqnarray}
the necessary and sufficient condition is
\begin{eqnarray}\label{dcc2}
\delta({\bf E}\hat\de g)=0.
\end{eqnarray}

\vskip 4mm {\bf 3.}  Let us now consider the
vacuum gravitational fields in GR. The
Hilbert-Einstein action in natural units  on
$({\cal M}^4, g)$ reads
\begin{equation}
{\bf L}=\frac{1}{16\pi}R{{\mbox {\boldmath
$\epsilon$}}}
\end{equation}
where $R$ is the scaler curvature and ${{\mbox
{\boldmath $\epsilon$}}}$ the volume element
determined by $g_{ab}$.


According to eq.(\ref{f}), the horizontal
variation of the Hilbert-Einstein action induced
by $f_{\lambda}$ can be calculated
\begin{eqnarray}
\hat\delta{\bf L}&=&
      \frac{1}{16\pi}[G_{ab}{\cal
      L}_{\xi}g^{ab}+\del_a(\Theta^a
      (\hat\delta g))]
{\mbox {\boldmath $\eps$}},
\end{eqnarray}
where $G_{ab}$ is the Einstein tensor and
$$\Theta^a(\hat\delta
g)=\frac{1}{16\pi}[2\nabla^a\nabla_b\xi^b-\nabla_b\nabla^a\xi^b-\nabla_b\nabla^b\xi^a
].$$
 On the
other hand, the diffeomophism invariance of the
Lagrangian directly leads to
\begin{eqnarray}
\hat\de\dl =\frac{1}{16\pi}{\cal
L}_{\xi}(R{{\mbox {\boldmath $\epsilon$}}})
=\frac{1}{16\pi}\del_c(R\xi^c){{\mbox {\boldmath $\epsilon$}}}.
\end{eqnarray}
From the above two equations, it follows the
covariant conservation law for the Noether
current eq.(\ref{cc}) with respect to the
diffeomophism invariance for the vacuum
gravitational fields in GR
\begin{eqnarray}\label{ccgr}
0=\frac{1}{16\pi}{G_{ab}}\hat\de g^{ab}+{\nabla^a
{\bf j}_a},
\end{eqnarray}
where the Noether current is  given by %
\begin{eqnarray}\label{cc1}
{\bf
j}_{a}=\frac{1}{8\pi}G_{ab}\xi^b+\frac{1}{16\pi}[\nabla_b\nabla^a\xi^b-\nabla_b\nabla^b\xi^a].
\end{eqnarray}

In this letter, we focus on the case of
stationary, axisymmeric  black holes and $ \xi$
being the Killing vector
\begin{eqnarray}\label{kv}
\xi^a=t^a+\Omega_H\phi^a, \end{eqnarray} where
$t^a$ and $\phi^a$ is the time-like and
space-like Killing vector of the space-time,
respectively. Since $\xi$ is a Killing vector,
the corresponding conserved current ${\bf j}$ is
given by eq. (\ref{cck}). In GR, it is the
current
\begin{eqnarray}\label{cckgr}
{\bf
j}_a
=\frac{1}{8\pi}R_{ab}\xi^b+\frac{1}{8\pi}G_{ab}\xi^b,
\end{eqnarray}
where the following equation has been used
\begin{eqnarray}\label{rk}
R_{eb}\xi^b=\frac{1}{4} \epsilon_{efab}\del^f
\epsilon^{abcd}\del_c\xi_d.
\end{eqnarray}

It is now very obvious but important to note that
this Noether current and consequently its charge
on a Cauchy surface $\Sigma$,
\begin{eqnarray}\label{NC}
{Q:=\int_{\Sigma} *{\bf j},}
\end{eqnarray}
{definitely vanish for the vacuum gravitational
fields with the Killing vector (\ref{kv}) due to
the }{Einstein equation. %
It is significant to emphasize, however,  that
although this Noether 
charge}{vanishes, it still leads to the mass
formula.}

Let us now show how this  vanishing Noether
charge leads to the  mass formula. From the
definition (\ref{NC}), this vanishing Noether
charge is given by
\begin{eqnarray}
{0\equiv Q=\frac{1}{8\pi}\int_{\Sigma} [\del ^b
\del _{[a} \xi_{b]}
    +G_{ab}\xi^b]{d \sigma}^a,}
\end{eqnarray}
where $d\sigma^{a}$ is the surface element on
$\Sigma$. {The first term is a total divergence,
whose integral} {may reduce to the surface one on
the boundary of $\Sigma$.  The second integral
vanishes due to } {the Einstein equation. Note
that the Cauchy surface $\Sigma$ has two
boundaries. One is at the }{spatial infinity
$S_{\infty}$ and the other is at the event
horizon $S_H^{(-)}$, where the upper index $(-)$}
{denotes the opposite orientation.} Thus, the
vanishing Noether charge is composed of two parts
${\cal Q}_{H}$ and ${\cal Q}_{\infty}$ at
$S_H^{(-)}$ and
$S_\infty$, respectively. Namely,
\begin{eqnarray}
Q&:=&{\cal Q}_{\infty}-{\cal Q}_{H}\nonumber\\
&=&\frac{1}{8\pi}\int_{S_{\infty}}%
\epsilon_{abcd}\del^c\xi^d d\sigma^{ab}
    -\frac{1}{8\pi}\int_{S_H}%
    \epsilon_{abcd}\del^c\xi^d
    d\sigma^{ab},
\end{eqnarray}
where $d\sigma^{ab}$ is the surface element on
$\partial\Sigma$.

By definition, 
the Komar mass and the angular momentum
are given by
\begin{eqnarray}
M&:=&\frac{1}{8\pi}\int_{S_{\infty}}\epsilon_{abcd}
\nabla^{c}t^{d}d\sigma^{ab},\nonumber\\
J&:=&-\frac{1}{16\pi}\int_{S_{\infty}}\epsilon_{abcd}
\nabla^{c}\phi^{d}d\sigma^{ab},
\end{eqnarray}
respectively. For the case of Killing vector, the
Komar mass $M$ is the same as the ADM mass of the
black hole configurations. Thus it follows the
expression for ${\cal Q}_\infty$
\begin{eqnarray}
{\cal Q}_\infty:=\frac{1}{8\pi}\int_{S_{\infty}}%
    \epsilon_{abcd}\del^c\xi^d
    d\sigma^{ab}=M-2\Omega_HJ.
\end{eqnarray}

On the other hand, %
for the Killing vector (\ref{kv}),
\begin{eqnarray}
\del^a\xi^b = \kappa \eps^{ab}
\end{eqnarray}
on $S_H$, where $\kappa$ is the surface gravity
and $\epsilon^{ab}$ the bi-normal to $S_H$.  Thus
it is easy to get the expression for ${\cal Q}_H$
\begin{eqnarray}
{\cal Q}_H:=\frac{1}{8\pi}\int_{S_H}%
\epsilon_{abcd}\del^c\xi^d d\sigma^{ab}
=\frac{\kappa}{8\pi}A,
\end{eqnarray}
where $A$ is the area of the event horizon. {It
is ${\cal Q}_H$ that is called as the Noether
charge in \cite{wald2}}.

Therefore, it has been shown that the vanishing
Noether charge, as a whole, with respect to the
diffeomorphism invariance leads to the mass
formula for the vacuum gravitational fields of
stationary, axisymmetric black holes in GR
\cite{jb}, \cite{sm}:
\begin{equation}\label{mass}
Q=M-2\Omega_HJ-\frac{\kappa}{8\pi}A=0.
\end{equation}

\vskip 4mm {\bf 4.} Let us now derive, from the
variation relation of this vanishing Noether
charge, the differential formula for mass i.e.
the first law of black hole thermodynamics, for
the vacuum stationary, axisymmetric black holes
in GR by the variational approach.

For the vacuum gravitational fields in GR,  the
variation relation of this vanishing Noether
current eq. (\ref{dcc}) becomes
\begin{eqnarray}\label{dccgr}
0=\frac{1}{16\pi}\delta({G_{ab}}\hat\de g^{ab})+
\delta[{\nabla^a {\bf j}_a}],
\end{eqnarray}
where ${\bf j}_a$ is given by (\ref{cc1}). As was
mentioned before, it is obvious  that the
conserved current $\bf j$ vanishes for the
Killing vector field $\xi^a$ (\ref{kv})  if the
vacuum Einstein equation holds. Further, if the
variation or the perturbation $\delta g$ is
restricted in such a way that both $g$ and
$g+\delta g$ are stationary, axisymmetric black
hole configurations and $\xi^a$ is the Killing
vector (\ref{kv}), the vanishing Nother current
now is (\ref{cckgr}) and the variation of the
conserved current should also vanish, i.e.
$\delta {\bf j}=0$ as well.

 Thus it is straightforward to get
\begin{eqnarray}\label{bch}
0&=&\frac{1}{8\pi}\int_{\Sigma} \de[R_{ab}\xi^b{d \sigma}^a]\nonumber\\
&=&\de[M-\frac{\kappa}{8\pi}A-2\Omega_HJ],
\end{eqnarray}
and  eq. (\ref{dcc2}) is also satisfied. Here
$\Sigma$ is the Cauchy surface.

 It should be noticed that  eq. (\ref{bch}) is just the start
point of Bardeen, Carter and Hawking's
calculation for the first law of the black hole
mechanics \cite{hawking}. As was required in
\cite{hawking}, under above perturbations, the
positions of event horizon and the two Killing
vector fields in (\ref{kv}) are unchanged.
Consequently, eq. (\ref{bch}) definitely gives
rise to the differential formula for mass, i.e.
the first law of black hole thermodynamics, among
the stationary, axisymmetric black hole
configurations \cite{jb}, \cite{sm}:
\begin{eqnarray}\label{dmass}
\delta M-\frac{\kappa}{8\pi}\delta
A-\Omega_H\delta J=0.
\end{eqnarray}

Thus both the mass formula (\ref{mass}) and its
differential formula (\ref{dmass}), i.e. the
first law of black hole thermodynamics for the
stationary, axisymmetric black hole
configurations are all derived from the
diffeomophism invariance of the Lagrangian by the
variational approach. Especially, they are in
fact the vanishing Noether charge and its
perturbation among the stationary, axisymmetric
black hole configurations.

\vskip 4mm {\bf 5.} Finally, a few remarks are in
order:

1. From the derivation via the variational
approach in this letter, it can be seen that
neither the entropy of black holes nor the total
energy of the gravitational fields is the
{entire} Noether charge of the conserved current
with respect to the diffeomorphism invariance {on
the Cauchy surface} at least for the vacuum
gravitational fields. In fact, the entire mass
formula  (\ref{mass}) itself, as a whole, should
be viewed as a Noether charge of the current
{eq.(\ref{cc1})} {on the Cauchy surface} and this
charge definitely vanishes.

2. At first glance it seems to be intricate why
the conserved current ${\bf j}$ (\ref{cckgr}) for
the stationary, axisymmetric black hole
configurations  vanishes. In fact, the Bianchi
identity may shed light on this point from
another angle of view, since
 the Bianchi identity may be
viewed as a consequence  of the diffeomophism
invariance. Using the Bianchi identity and the
vacuum Einstein equation, it is easy to shown
\begin{eqnarray}
\del^a(G_{ab}\xi^b)=0,\quad \forall \xi. 
\end{eqnarray}
This means that the conserved current ${\bf j}$
can be gotten from the Bianchi identity up to a
constant factor. Therefore, at least for the
stationary, axisymmetric black holes, it could
also be shown that the mass formula and the first
law of black hole thermodynamics are the
consequences of the Bianchi identity and its
perturbation as well. In addition, this also
indicates why the Noether charge of the conserved
current ${\bf j}$  should vanish.

On the other hand, it also seems to be a
consequence of the equivalence between
gravitational mass and inertial mass. In ordinary
local field theories including the gravitational
field, the charge, as an integral over a Cauchy
surface of the time-like component of the
conserved current with respect to the
diffeomorphism invariance under some orthogonal
normal tetrad, is the local energy density. But
the above equivalence indicates that there should
be no local energy density for the gravitational
field in general relativity. Otherwise, if there
were local mass for the gravitational fields, not
only this equivalence could no longer be correct
but GR even Newton's theory could be reformulated
to fit the observations. The vanishing Noether
current and its vanishing charge with respect to
the diffeomorphism invariance just reflect this
fundamental property for the gravitational
fields.

3. It should be noticed that the requirement on
$\xi^a$ being a Killing vector might  not be
necessary. In fact, for the vector $\xi^a$ what
are needed possibly  its asymptotic behavior near
the horizon and that at the spatial infinity. If
it is required some appropriate quasi-local
horizon condition \cite{ash} such as isolated
horizon, the mass formula  (\ref{mass}) could
also be gotten. In other wards, the derivation of
the mass formula might be generalized into some
non-stationary space-time configurations, which
have been considered by other authors from
different motivation \cite{ash}.

4. In this letter what have been dealt with are
the vacuum gravitational fields of 4-dimensions.
In principle, all these results may be formulated
for the gravitational fields with sources and of
other dimensions. This topic will be published
elsewhere.

\vskip 6mm

\centerline{\bf Acknowledgments} \vskip 2mm We
would like to thank Prof. R.M. Wald for valuable
discussions on this issue during his visit to
ITP, Chinese Academy of Sciences on August %
2001. This work was supported in part by the
National Natural Science Foundation of China
grant Nos. 90103004, 10175070  and the National
Key Project for Basic Research of China
(G1998030601).


\begin{thebibliography}{99}
\bibitem{wald1} Wald R.M., The First Law of Black Hole Mechanics,
in {\it Directions in General Relativity}, vol.
1, ed. by B.L. Hu, M. Ryan, and C.V. Vishveshwara
Cambridge Univ.Press, Cambridge, 1993.
gr-qc/9305022.
\bibitem{wald2} Wald R.M., Black Hole Entropy Is the Noether
Charge, Phys. Rev. {\bf D48} (1993) R3427-R3431.
gr-qc/9307038.
\bibitem{wald3}Iyer  V. and Wald R.M.,  Some Properties of Noether Charge and a Proposal for
Dynamical Black Hole Entropy, Phys. Rev. {\bf
D50} (1994) 846-864. gr-qc/9403028.
\bibitem{wn} Nelson W., A Comment on Black Hole Entropy in String
Theory, Phys. Rev. {\bf D50} (1994) 7400-7402.
hep-th/9406011.
\bibitem{wald5} Wald R.M., The Thermodynamics of Black Holes,
Living Rev. Rel. {\bf 4} (2001) 6. gr-qc/9912119.
and references therein.
\bibitem{jb} Bekenstein J.D., Black Holes and
Entropy, Phys. Rev. {\bf D7} (1973) 2333-2346.
\bibitem{sm} Smarr L., Mass Formula for Kerr Black
Holes, Phys. Rev. Lett. {\bf 30}, (1973) 71-73.
\bibitem{hawking} Bardeen J.M., Carter B. and Hawking S.W., The Four Laws of
Black Hole Mechanics, Commun. Math. Phys. {\bf
31} (1973) 161-170.

\bibitem{ash}Ashtekar A., Beelte C. and Fairhurst S., Mechanics of Isolated Horizons,
Class. Quant. Grav. {\bf 17} (2000) 253-298 and
references therein. gr-qc/9907068.

\end{thebibliography}
\end{document}